\documentclass[aps,prb]{revtex4}
\usepackage{amsmath}
\usepackage{graphicx}

\begin{document}

\title{  Enhancement of Fluctuation-Induced Electromagnetic Phenomena in dynamically nonequilibrium systems at Resonant Photon Emission}

\author{A.I.Volokitin\footnote{
\textit{E-mail address}:alevolokitin@yandex.ru}}
 
\affiliation{
Samara State Technical University,  443100 Samara, Russia}

\begin{abstract}

We study    the resonances  in   Casimir friction, radiative heat transfer and heat generation for two plates sliding relative to each other. Resonances have a different origin  in the frequency range of the \textit{normal} (NDE) and \textit{anomalous} (ADE) Doppler effect. In the frequency range of NDE,   resonances are associated with  resonant photon tunnelling between  surface phonon/plasmon polaritons of  plates. For two identical plates, such  resonances exist only at a relative sliding velocity $v=0$. However, for different plates such resonance can  exist at $v\neq 0$. In the frequency range of ADE,  resonances  are associated  with   the creation of  excitations in  both plates. While  in the frequency range of NDE the photon emission rate has an upper limit, in the frequency range of ADE, the photon emission rate can diverge  even in the presence of  dissipation in the system. We consider resonances  for the  identical and different sliding plates. We discuss the possibility to detect Casimir friction with its limiting case of quantum friction using an atomic force microscope in graphene structures.

\end{abstract}
\maketitle


\vskip 5mm

\section{Introduction}

Quantum  fluctuations of the electromagnetic field manifest themselves in a wide variety of fields of physics. 
For example, the Lamb shift of atomic spectrum and anomalous magnetic moment of the electron were explained with the help of this idea.  
Most directly, these fluctuations are manifested through the Casimir effect. In the late 1940s Hendrik Casimir predicted \cite{Casimir1948} that two macroscopic non-magnetic bodies 
with no net electric charge (or charge moments) can experience an attractive force much stronger than gravity. The existence of this force is one of the direct 
macroscopic manifestations of quantum mechanics.
Hendrik Casimir based his prediction on a simplified model involving two parallel perfectly conducting plates separated by vacuum. 
A unified theory of both the van der Waals and Casimir forces between plane parallel material plates, in thermal equilibrium and separated by a vacuum gap, was developed by Lifshitz (1955) 
\cite{Lifshitz1955}. To calculate the interaction force Lifshitz used Rytov’s theory of the fluctuating electromagnetic field \cite{Rytov1989}.  
At present the interest of Casimir forces is increasing because they 
dominate the interaction between nanostructures, and are offen responsible for the stiction  between moving parts in small devices such as micro- and nanoelectromechanical systems, and can be 
considered as practical mechanisms for the actuation of such devices.  Due to this practical interest, and the fast progress in force detection techniques, experimental and theoretical
 investigations of Casimir forces have experienced an extraordinary “renaissance” in the past few years (see \cite{Dalvit2011} for a review and the references therein).

\begin{figure}[tbp]
\includegraphics[width=0.40\textwidth]{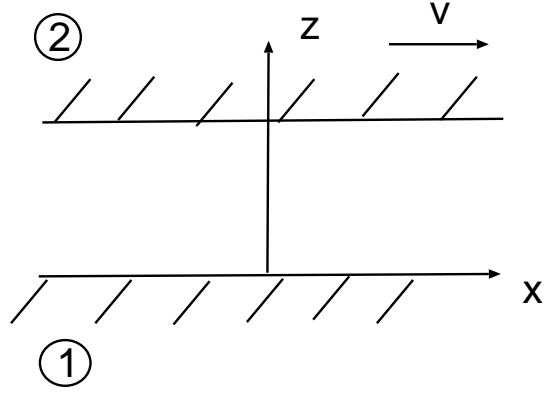}
\caption{Two semi-infinite bodies with plane parallel surfaces separated by
a distance $d$. The upper solids moves parallel to other with the velocity $v$. }
\label{Fig1}
\end{figure}

Another manifestation of  quantum fluctuations of the electromagnetic field is the noncontact quantum friction between bodies in 
relative motion \cite{Fundamentals2015,VolokitinRMP2007,VolokitinBook2017}. The noncontact friction determines the ultimate limit to which the friction force 
can be reduced and, consequently, also the force fluctuations because they are linked to friction  via the fluctuation-dissipation theorem. 
The force fluctuations (and hence friction) are important for the ultrasensitive 
force detection.
Friction is usually a very complicated process. The simplest case consists of two flat surfaces, separated by a vacuum gap (see Fig.(\ref{Fig1}),  sliding relative to each other at $T=0$ K, where the friction is generated by the relative movement of quantum  fluctuations\cite{PendryJPCM1997,PendryJMO1998,VolokitinPRL2003,VolokitinPRB2003,
VolokitinJPCM1999,VolokitinRMP2007,VolokitinPRB2006,VolokitinPRB2008,
VolokitinPRL2011,VolokitinBook2017}.    The thermal and quantum fluctuations of the current density in one body induces 
the current density in other body; the interaction between these current densities 
is the origin of the Casimir interaction. When two bodies are in relative motion, 
the induced current will lag slightly behind the fluctuating current inducing it, and 
this is the origin of the Casimir friction.  At present the Casimir friction is attracting a lot of attention due to the fact that it is one of the mechanisms of noncontact friction between bodies in the absence of direct contact \cite{VolokitinRMP2007,Fundamentals2015,VolokitinBook2017}. 
The Casimir friction  was studied in the configurations plate-plate \cite{VolokitinRMP2007,VolokitinBook2017,PendryJPCM1997,PendryJMO1998,
VolokitinJPCM1999,VolokitinPRL2003,VolokitinPRB2003,VolokitinPRB2006,
VolokitinPRB2008,
VolokitinPRL2011,VolokitinPRB2016a},  neutral particle-plate \cite{VolokitinRMP2007,VolokitinBook2017,TomassonePRB1997,VolokitinPRB2002,
VolokitinPRB2006,DedkovPLA2005,DedkovJPCM2008,BartonNJP2010,BrevikEntropy2013,BrevikEPJD2014,KardarPRD2013,
DalvitPRA2014,VolokitinNJP2014,HenkelNJP2013,HenkelJPCM2015,VolokitinJETPLett2016},  and neutral particle-blackbody radiation  \cite{VolokitinRMP2007,VolokitinBook2017,VolokitinPRB2008,HenkelNJP2013,MkrtchianPRL2003,
DedkovNIMPR2010,
JentschuraPRL2012,JentschuraPRL2015,VolokitinPRA2015}. While the predictions of the theory for the Casimir forces were verified in many experiments \cite{Dalvit2011}, the detection of the 
Casimir friction  is still challenging problem for  experimentalists. However, the frictional  drag between quantum wells \cite{GramilaPRL1991,SivanPRL1992} and  graphene sheets 
\cite{KimPRB2011,GeimNaturePhys2012}, and the current-voltage dependence of nonsuspended graphene on the surface of the polar dielectric SiO$_2$ \cite{FreitagNanoLett2009}, were accurately described using the 
theory of the Casimir friction \cite{VolokitinPRL2011,VolokitinJPCM2001b,VolokitinEPL2013}. At present the frictional drag experiments \cite{GramilaPRL1991,SivanPRL1992,KimPRB2011,GeimNaturePhys2012,FreitagNanoLett2009} were performed for weak electric field when the induced drift motion of the free carriers is smaller than 
the threshold velocity for quantum friction. Thus in these experiments the frictional drag is dominated by the contributions from  thermal fluctuations. However, the measurements of the current-
voltage dependence \cite{FreitagNanoLett2009} were performed for high electric field, where the drift velocity is above the threshold velocity, and where the frictional drag is dominated by  quantum fluctuations \cite{VolokitinPRL2011,VolokitinEPL2013}. 

Quantum friction is associated with creation of  excitations of  different kind. 
For transparent dielectrics such excitations are photons and quantum friction 
is associated with quantum Vavilov-Cherenkov radiation.  Thus there is a close connection between quantum 
friction and the quantum Vavilov-Cherenkov 
radiation -- both of these phenomena are related with the instability of the quantum state against spontaneous production of elementary excitations in the conditions of the  anomalous Doppler effect 
\cite{Frank1943,Ginzburg1945, Ginzburg1996,PendryJPCM1997,VolokitinRMP2007,VolokitinPRL2011,KardarPRA2013,SilveirinhaPRX2014,VolokitinPRB2016a}. According to Landau \cite{Landau1941} the same instability breaks superfluid order beyond a certain velocity. Quantum  Vavilov-Cherenkov radiation was first  described by  Frank \cite{Frank1943} and   Ginzburg and  Frank \cite{Ginzburg1945} (see also \cite{Tamm1959,Ginzburg1996} for 
review of these work). Quantum friction is associated with creation of surface phonon polaritons for  dielectrics or     
 surface plasmon polaritons for metals. If an object has no internal degrees of freedom (e.g., a point charge), then the energy of the radiation is determined by the change of the  kinetic energy of the object. However, if an object has  internal degrees of freedom (say, an atom), then two types  of excitations 
 are possible.  If the frequency of the excitation in the \textit{lab} reference frame $\omega >0$, then in  the rest frame of an object, due to the Doppler effect, 
the frequency of the radiation $\omega^{\prime}=\omega-k_xv$. In the normal Doppler effect frequency region, 
when $\omega^{\prime}>0$, the excitation energy is determined by the decrease of the internal energy. For example, for an atom the state may changes from the excited state $|1>$ to the 
ground state 
$|0>$. The region of the \textit{anomalous} Doppler effect corresponds to $\omega^{\prime}<0$ in which case an object becomes excited when it creates an excitation in the \textit{lab} reference frame. For example, an atom could  experience the transition from the ground state $|0>$ to the excited state $|1>$ when it radiates. In such a case energy conservation requires that the energy  of   the excitations result from a
 decrease of the kinetic energy of the object. That is, the self-excitation of a system is accompanied by a slowing down of the motion of the object as a whole. 
For a neutral object the interaction of the object 
with the matter is determined by the fluctuating electromagnetic field due to the quantum fluctuations inside the object. 

 While a constant translational motion  requires for the emission of the radiation at least two bodies in relative motion (otherwise it is not possible due to Lorenz invariance), a single accelerated object 
can radiate and experience friction. Quantum fluctuations of the electromagnetic field are determined by virtual photons that are continuously created and annihilated in the vacuum. Using a metal mirror in   
accelerated motion, with velocities  near the light velocity, virtual photons can be converted into real photons, leading to radiation emitted by the mirror. This is the dynamic Casimir effect \cite{Moore1970,Danies1976,Dalvit2011}; which recently 
 was observed in a superconducting waveguide \cite{Wilson2011}. Radiation can be also emitted by a rotating object \cite{ManjavacasPRL2010,ManjavacasPRA2010,KardarPRL2012,KardarPRA2014,BercegolPRL2015}. In fact this phenomenon is closely related to the  prediction 
by Zel'dovich \cite{ZeldovichJETPLett1971} of  an amplification of  
certain waves during scattering from a rotating body. Rotation quantum friction  acting on a  nanoparticle rotating  near the surface was studied in 
\cite{PendryPRL2012,DedkovEPL2012,ManjavacasPRL2017,DedkovTekhPhys2017,VolokitinEPL2018,VolokitinJETPLett2018}. The authors of \cite{PendryPRL2012} showed that quantum friction 
increases strongly at resonant generation of surface phonon polaritons at the rotation frequency $\Omega=\omega_1+\omega_2$, where $\omega_1$ and $\omega_2$ are the frequencies of 
surface polaritons for the particle and surface, respectively. In \cite{VolokitinEPL2018,VolokitinJETPLett2018}  it was shown that this system allows 
singular resonance at which the electromagnetic field increases unlimitedly with time. Stationary rotation is impossible at resonance because the friction torque increases unlimitedly with time. 
However, stationary rotation is possible near resonance. In this case, fluctuation-induced electromagnetic effects increase strongly. Singular resonance at the relative sliding of two identical plates was 
first predicted in \cite{JacobJOpt2014,JacobOptExpress2014} and was then studied in \cite{VolokitinPRB2016}. Singular resonance for two rotating nanoparticles was 
first predicted in \cite{VolokitinJETPLett2017,VolokitinPRA2017}. Singular resonance for a body rotating in vacuum is impossible because it requires at least two bodies. 
However, singular resonance is possible for a body rotating in a cavity. This effect was first predicted by Zel’dovich \cite{ZeldovichJETPLett1970,ZeldovichJETP1972} and 
was recently studied for a cavity with rotating walls \cite{SilveirinhaPRA2016}. Resonant photon tunnelling enhancement of the radiative heat transfer between two plates at rest was considered in Ref. \cite{VolokitinPRB2004}. In this article more general study  of the resonances which exist in Casimir friction, radiative heat transfer and heat generation for two plates sliding relative to each other is presented. We consider the resonances which exist in the frequency range of the \textit{normal} and \textit{anomalous} Doppler effect. 

\section{Casimir friction between two plates \label{PP}}

\subsection{General theory}

We consider two plates sliding relative to each other with velocity $v$ and separated by a distance $d$ (see Fig.\ref{Fig1}). According to the theory of the Casimir friction \cite{VolokitinJPCM1999,VolokitinRMP2007,VolokitinPRB2008,VolokitinPRL2011} the contributions  of the evanescent waves (which dominate for short separation between plates,  large velocities and low temperatures)  to the friction force 
$F_{1x}$, and the heat power  $P_1$ absorbed by plate  \textbf{1} in the nonrelativistic and nonretarded limit in the    \textit{lab} reference frame, are determined by formulas

\begin{equation}
\left(\begin{array}{c}
F_{1x}\\
P_1
\end{array} \right)
 = \int \frac{d^2q}{(2\pi)^2}
\int_0^{\infty} \frac{d\omega}{2\pi} \left(\begin{array}{c}
\hbar q_x\\
\hbar \omega
\end{array} \right)\Gamma_{12}(\omega, q)\mathrm{sgn}(\omega^-)[n_2(\omega^-) - n_1(\omega))]
\label{qvc1}
\end{equation}
where the positive quantity
\begin{equation}
\Gamma_{12}(\omega, \mathbf{q})= 4\mathrm{sgn}(\omega^-) \left\{\frac{\mathrm{Im}R_{1p}\mathrm{Im}R_{2p}^-}{|\Delta_{pp}|^2}
 +(p\leftrightarrow s)\right\}e^{-2 q d}
\end{equation}
can be identified as a spectrally resolved photon emission rate,
\[
\Delta_{pp} = 1 - e^{-2qd}R_{1p}R_{2p}^-,
\]
$n_i(\omega)=[\exp(\hbar\omega/k_BT_i)-1]^{-1}$,  
$R_{1 p(s)}$ is the reflection amplitude for surface
\textbf{1} in the  rest reference frame for   a $p(s)$ - polarized electromagnetic wave, 
$R_{2 p(s)}^- = R_{2 p(s)}(\omega^-, q)$ is the reflection 
amplitude for surface
\textbf{2}  in the rest frame of plate \textbf{2} for   a $p(s)$ - polarized electromagnetic wave, 
$\omega^-=\omega-q_xv$. 
The symbol
$(p\leftrightarrow s$) denotes the terms that are obtained from the preceding terms by 
permutation of indexes   $p$ and $s$. In the domains of the \textit{normal} 
Doppler effect  ($\omega - q_xv >0$) the last factor in the integrand in Eq. 
(\ref{qvc1}) can be written in the form
\[
\mathrm{sgn}(\omega^-)[n_2(\omega^-) - n_1(\omega)] = n_2(\omega^-)[1+n_1(\omega))] - n_1(\omega)[1+n_2(\omega^-)].
\]
Thus, in this domain  the energy and momentum transfer are related as when the excitations are annihilated in one body and  created in other body. Such processes are only possible at  $T \ne 0$ K, i.e. they are associated with thermal radiation.  On the other hand, in the case of the \textit{anomalous} Doppler effect   ($\omega - q_xv <0$)
\[
\mathrm{sgn}(\omega^-)[n_2(\omega^-) - n_1(\omega))] = 1 + n_2(|\omega^-|)+n_1(\omega) = [1 + n_2(|\omega^-|)][1 + n_1(\omega)] -n_2(|\omega^-|)n_1(\omega).
\]
In this case the excitations are created and annihilated  simultaneously in both bodies. 
Such processes are possible even at  $T = 0$ K and are associated with quantum friction.

At  $
T_1=T_2=0$ K the propagating waves do not contribute to the friction and the radiative 
heat transfer. However, the contribution from the evanescent waves does not vanish. 
Taking into account that  $n_1(\omega )=0$ at $
T_1=T_2=0$ K and $\omega >0$, 
\[
n_2(\omega^-)=\left\{
\begin{array}{rl}
-1 & \text{for} \,\,0<\omega < q_xv \\
0& \text{for} \,\,\omega > q_xv
\end{array}\right.
\]
 Eq.  (\ref{qvc1}) is reduced to  the formula    obtained by 
Pendry in Ref. \cite{PendryJPCM1997}
\begin{equation}
\left(
\begin{array}{c}
F_x\\
P_1
\end{array} \right)
=
-\frac {\hbar} {\pi ^3}\int_0^\infty dq_y \int_0^\infty dq_x
\int_0^{q_xv} d\omega \left(\begin{array}{c}
\hbar q_x\\
\hbar \omega
\end{array} \right)
\left(\frac{\mathrm{Im}R_{1p}\mathrm{Im}R_{2p}^{-}}{|D_{pp}|^2} + 
\frac{\mathrm{Im}R_{1s}\mathrm{Im}R_{2s}^{-}}{|D_{ss}|^2}\right)e^{-2 q d},
\label{approximate}
\end{equation}

\subsection{Resonances in the Casimir friction}

\subsubsection{The frequency range of the normal Doppler effect}

In the frequency range  of the \textit{normal} Doppler effect, when $\omega>0$
 and $\omega-\Omega > 0$,   we can write  $R_{1p}(\omega)=|R_{1p}(\omega)|\mathrm{exp}(i\phi_i)$ and $R_2(\omega-q_xv)= |R_{2p}(\omega-q_xv)|\mathrm{exp}(i\phi_2)$, and the photon emission rate   can be written in the form
\[
\Gamma_{12}=\frac{4\mathrm{Im}R_1(\omega)\mathrm{Im}R_2(\omega-q_xv)e^{-2qd}
}
{|1-e^{-2qd}R_p(\omega)R_2(\omega-q_xv))|^2}=
\]
\begin{equation}
\frac {4|R_1(\omega)||R_2(\omega-q_xv)|e^{-2qd}\mathrm{sin}\phi_1\mathrm{sin}\phi_2}{1+|R_1(\omega)|^2|R_2(\omega-q_xv)e^{-2qd}|^2-2|R_1(\omega)||R_2(\omega-q_xv)e^{-2qd}|
\mathrm{cos}(\phi_1+\phi_2)},
\label{nomDoppler}
\end{equation}
which is maximal ($\Gamma_{max}=1$) for $|R_1(\omega)||R_2(\omega-q_xv)e^{-2qd}|=1$ and $\phi_1=\phi_2$. Thus at $v=0$ $P_1\leq P_{max}$, where
\begin{equation}
P_{max}=\frac{ k_B^2}{48\hbar}\left(T_2^2 - T_1^2\right)q_c^2
\end{equation}
where $q_c$ is a cut of in $q$, determined by the properties of the materials. The largest possible $q_c\sim 1/b$ where $b$ is an inter-atomic distance. Thus the ration of the maximal heat flux connected with evanescent waves to heat flux due to black body radiation $P_{max}/P_{BB}\sim (\lambda_T/b)^2$ where $\lambda_T=c\hbar/k_BT$. At room temperature, the maximal contribution to the heat flux from the evanescent waves will be approximately in 10$^8$ times larger than the contribution from black body radiation. 

The radiative heat transfer   between two plates is strongly enhanced in the case of the resonant photon tunneling  \cite{VolokitinRMP2007,VolokitinBook2017,VolokitinPRB2004}. The reflection amplitude for the dielectric plate for $d<c/(\omega_i|\epsilon_i|)$ 
\begin{equation}
R_{ip}=\frac{\varepsilon _{i}-1}{\varepsilon _{i}+1},
 \label{refcoef}
\end{equation}
where $\varepsilon _{i}$ and $\omega_i$ are  the dielectric function and the phonon polariton frequency for dielectric \textit{i}. The  dielectric surfaces have resonances at  $\varepsilon_i^{\prime}(\omega_i)=-1$ where $\varepsilon_i^{\prime}$
 is the real part of $\varepsilon_i$. For a polar dielectric $\omega_i$  determines the frequency of the surface phonon polariton. Near the resonance at $\omega \approx
\omega_1$ and $\omega-q_xv\approx \omega_2$ for $2/\varepsilon_i^{\prime\prime}>>1$ in the frequency region of the normal Doppler effect the reflection amplitudes
for  dielectrics can be written in the form
\begin{equation}
R_{1p}(\omega)\approx -\frac{a_1}{\omega - \omega_1 +i\Gamma_1},\,\,\,
R_{2p}(\omega-q_xv)\approx -\frac{a_2}{\omega-q_xv-  \omega_2 +i\Gamma_1},
\label{res.nom}
\end{equation}
where
\begin{equation}
a_i=\frac {2}{(d/d\omega)\varepsilon_i^{\prime}(\omega)|_{\omega=\omega_i}},\,\,\,
\Gamma_i = \frac {\varepsilon_i^{\prime \prime}(\omega_i)}{(d/d\omega)
\varepsilon_i^{\prime}(\omega)|_{\omega=\omega_i}},
\end{equation}

For two identical surfaces, when $\omega_1=\omega_2=\omega_0$, $a_1=a_2=a$ and $\Gamma_1=\Gamma_2=\Gamma$, the resonant condition $\phi_1=\phi_2$ can be only satisfied at $v=0$. The second resonance condition $|R_p(\omega)|^2e^{-2qd}=1$ is satisfied  at $\omega=\omega_{\pm}$ where
\begin{equation}
\omega_{\pm}=\omega_0\pm\sqrt{a^2e^{-2qd}-\Gamma^2}.
\end{equation}
Close to the resonance the photon emission  rate    for two identical dielectric plates  can be written in the form
\begin{equation}
\Gamma_{12}=\frac {4(\mathrm{(Im}R_de^{-qd})^2}{\left|
1-e^{-2qd}R_{p}^2\right| ^2}\approx \frac {4(a\Gamma e^{-qd})^2}{[(\omega-\tilde{\omega}_+)^2+\Gamma^2][(\omega-\tilde{\omega}_-)^2+\Gamma^2]}
\label{gamnd}
\end{equation}
where
\begin{equation}
\tilde{\omega}_{\pm}=\omega_0\pm ae^{-qd}.
\end{equation}
Thus $\Gamma_{12}= 1$ at $\omega=\omega_{\pm}$. Using Eq. (\ref{gamnd}) in Eq. (\ref{qvc1})  gives the resonant contribution to the heat transfer  
\[
P_1 \approx \frac {\hbar\omega_0\Gamma} {2\pi } 
\int_0^\infty dq\,q\frac {[2e^{-qd}/\varepsilon^{\prime\prime}(\omega_0)]^2}{[2e^{-qd}/\varepsilon^{\prime\prime}(\omega_0)]^2+1} \approx \frac {\hbar\omega_0\Gamma} {2\pi } 
\int_0^{q_c} dq\,q
\]
\begin{equation}
=\frac {\hbar\omega_0\Gamma q_c^2} {4\pi }[n_2(\omega_0)-n_2(\omega_0)],
\label{resnd}
\end{equation}
where $\varepsilon^{\prime\prime}$ is the imaginary part of $\varepsilon$, $q_c=\mathrm{ln}[2/\varepsilon^{\prime\prime}(\omega_0)]/d$. It was assumed that $2/\varepsilon^{\prime \prime}(\omega_0)  \gg 1$, $\omega_0\gg a\mathrm{exp}(-qd)$. 

To linear order in the velocity $v$ the friction force $F=\gamma v$ where 
at $T_1=T_2=T$, the 
friction coefficient  
\begin{equation}
\gamma =\frac {\hbar^2} {8\pi ^2k_BT}\int_0^\infty
\frac{d\omega}{ \sinh^2\left(\frac{\hbar \omega}{2k_BT}\right)}
\int_0^\infty dq\,q^3e^{-2qd}\frac {\mathrm{Im}R_{1p}\mathrm{Im}R_{2p}}{\left|
1-e^{-2qd}R_{1p}R_{2p}\right| ^2}.
\label{parallel6}
\end{equation}
Using (\ref{gamnd}) in (\ref{parallel6})   gives the resonant contribution to the friction coefficient
\[
\gamma_{res} \approx \frac {\hbar^2\Gamma} {4\pi k_BT
 \sinh^2\left(\frac{\hbar \omega_0}{2k_BT}\right)}
\int_0^\infty dq\,q^3\frac {[2e^{-qd}/\varepsilon^{\prime\prime}(\omega_0)]^2}{[2e^{-qd}/\varepsilon^{\prime\prime}(\omega_0)]^2+1} \approx \frac {\hbar^2\Gamma} {4\pi k_BT
 \sinh^2\left(\frac{\hbar \omega_0}{2k_BT}\right)}
\int_0^{q_c} dq\,q^3
\]
\begin{equation}
=\frac {\hbar^2\Gamma q_c^4} {16\pi k_BT
 \sinh^2\left(\frac{\hbar \omega_0}{2k_BT}\right)},
\label{resfrcoeff}
\end{equation}
 At small frequencies far from the resonance ($\omega\ll\omega_0$) $\Gamma_{12}\approx (\omega/\omega^*))^2$ and the off-resonant contribution to the friction coefficient 
\begin{equation}
\gamma_{offres}\approx=\frac {\hbar} {16d^4}
\left(\frac{k_BT}{\hbar \omega^*}\right)^2.
\label{offres}
\end{equation}
For example, the dielectric function of amorphous SiO$_2$ can be described
using an oscillator model\cite{Chen2007APL}
\begin{equation}
\varepsilon(\omega) =
\epsilon_{\infty}+\sum_{j=1}^2\frac{\sigma_j}{\omega_{0,j}^2-\omega^2-i\omega\gamma_j},
\label{eps}
\end{equation}
where parameters $\omega_{0,j}$, $\gamma_j$ and $\sigma_j$ were
obtained by fitting the actual $\varepsilon$ for SiO$_2$ to the above
equation, and are given by $\epsilon_{\infty}=2.0014$,
$\sigma_1=4.4767\times10^{27}$s$^{-2}$, $\omega_{0,1}=8.6732\times
10^{13}$s$^{-1}$, $\gamma_1=3.3026\times 10^{12}$s$^{-1}$,
$\sigma_2=2.3584\times10^{28}$s$^{-2}$, $\omega_{0,2}=2.0219\times
10^{14}$s$^{-1}$, and $\gamma_2=8.3983\times 10^{12}$s$^{-1}$. 
Using these parameters in (\ref{eps}) gives  $\omega_0=9.29\cdot 10^{13}$s$^{-1}$, $a=3.6\cdot 10^{12}$s$^{-1}$, $\Gamma=1.8\cdot 10^{12}$s$^{-1}$, $\varepsilon^{\prime \prime}(\omega_0)=1$, $\omega^*=2.3\cdot 10^{16}$s$^{-1}$. With these 
parameters Eqs. (\ref{res}) and (\ref{offres}) at $T=300$K and $d=1$nm give $\gamma_{res}=3.5\cdot 10^{-2}$kgs$^{-1}$m$^{-2}$ and  $\gamma_{offres}=1.8\cdot 10^{-5}$kgs$^{-1}$m$^{-2}$.

For two different plates  resonance is possible when the real part of the reflection amplitudes  $R_{ip}^{\prime}(\omega_i)=0$ at the frequency of the surface phonon/plasmon polariton $\omega_i$. In this case $\phi_1=\phi_2=\pi/2$ for $q_xv=\omega_1-\omega_2$. The second condition for  resonance requires that at $q_x=(\omega_1-\omega_2)/v$  and $q_y=0$ 
\begin{equation}
R_{1p}^{\prime\prime}(\omega_1)R_{2p}^{\prime\prime}(\omega_2)\mathrm{exp}\left(-2\frac{|\omega_1-\omega_2|}{v}\right)=1
\label{cond.nom.res}
\end{equation}
where $R_{ip}^{\prime\prime}(\omega)$ is the imaginary part of the reflection amplitude. From (\ref{cond.nom.res}) follows that  resonance is possible for $v>v_c$ where
the critical velocity   
\begin{equation}
v_c=\frac{|\omega_1-\omega_2| d}{\mathrm{ln}[R_{1p}^{\prime\prime}(\omega_1)R_{2p}^{\prime\prime}(\omega_2)]^{1/2}}.
\label{cr.vel.norm}
\end{equation}
For the reflection amplitudes given by Eq. (\ref{res.nom}) $R_{ip}^{\prime\prime}(\omega_i)=a_i/\Gamma_i=2/\varepsilon_i^{\prime \prime}(\omega_i)$ and the critical velocity
\begin{equation}
v_c=\frac{|\omega_1-\omega_2| d}{\mathrm{ln}2/[\varepsilon_{1}^{\prime\prime}(\omega_1)\varepsilon_{2}^{\prime\prime}(\omega_2)]^{1/2}}
\end{equation}
For $\omega \approx \omega_1$ and $\omega-
q_xv\approx\omega_2$  the photon emission rate   can be written in the form
\begin{equation}
\Gamma_{12}\approx
\frac{4\Gamma_1\Gamma_2a_1a_2e^{-2qd}}
{(\Gamma_1+\Gamma_2)^2(\omega -\omega_c)^2
+\left[\Gamma_1\Gamma_2\left(\frac{(q_xv-\omega_1+\omega_2)}{\Gamma_1+\Gamma_2}
\right)^2
-(\omega -\omega_c)^2+ \frac{(\Omega-\omega_1+\omega_2)(\Gamma_2-\Gamma_1)(\omega-\omega_c)}
{\Gamma_1+\Gamma_2}+
\Gamma_1\Gamma_2+a_1a_2e^{-2qd}\right]^2}
\label{nom.res}
\end{equation}
where 
\begin{equation}
\omega_c=\frac{\Gamma_1(q_xv+\omega_2)+\Gamma_2\omega_1}{\Gamma_1+\Gamma_2}.
\end{equation}
The photon emission rate given by Eq. (\ref{nom.res}) has a maximum $\Gamma_{12}^{max}=1$ at $\omega = \omega_c$, $q_xv=\omega_1-\omega_2$ when the condition (\ref{cond.nom.res}) is fulfilled. Using (\ref{tantipart}) in (\ref{qvc1}) for $\omega -q_xv>0$ gives the resonant contributions at $v=v_c$ to  the  heat transfer rate and  friction force for $|\omega_1-\omega_2|>>\sqrt{\Gamma_1\Gamma_2}$ from the frequency range of the normal Doppler effect
\begin{equation}
\left(
\begin{array}{c}
F_x\\
P_1
\end{array} \right)
\approx
\frac{\Gamma_1\Gamma_2c^2}{\pi d^2|\omega_1-\omega_2|}\left(\begin{array}{c}
\hbar q_c\\
\hbar \omega_1
\end{array}\right) [n_2(\omega_2)-n(\omega_1)]
\label{nfhg}
\end{equation}
where $c=\mathrm{ln}2/[\varepsilon_{1}^{\prime\prime}(\omega_1)\varepsilon_{2}^{\prime\prime}(\omega_2)]^{1/2}$ and $q_c=c/d$.

\subsubsection{The frequency range of the anomalous Doppler effect}

In the frequency range  of the \textit{anomalous} Doppler effect, when $\omega>0$
 and $\omega-\Omega < 0$,   we can write  $R_{1p}(\omega)=|R_{1p}(\omega)|\mathrm{exp}(i\phi_1)$ and $R_2(\omega-q_xv)=R_2^*(q_xv-\omega) =|R_{2p}(\omega-q_xv)|\mathrm{exp}(-i\phi_2)$, and the photon emission rate   can be written in the form
\[
\Gamma_{12}=-\frac{4\mathrm{Im}R_1(\omega)\mathrm{Im}R_2(\omega-q_xv)e^{-2qd}
}
{|1-e^{-2qd}R_p(\omega)R_2(\omega-q_xv))|^2}=
\]
\begin{equation}
\frac {4|R_1(\omega)||R_2(\omega-q_xv)|e^{-2qd}\mathrm{sin}\phi_1\mathrm{sin}\phi_2}{1+|R_1(\omega)|^2|R_2(\omega-q_xv)e^{-2qd}|^2-2|R_1(\omega)||R_2(\omega-q_xv)e^{-2qd}|
\mathrm{cos}(\phi_1-\phi_2)},
\end{equation}
which diverges ($\Gamma_{12}=\infty$) for $|R_1(\omega)||R_2(\omega-q_xv)e^{-2qd}|=1$ and $\phi_1=\phi_2$. For two identical plates $\phi_1=\phi_2$ for $q_xv=2\omega$
or $q_x= 2\omega/v$ because in this case $R_2(\omega-q_xv)=R_1(-\omega)=R_1^*(\omega)$. The second condition for the occurrence of singular resonance at $q_x= 2\omega/v$ takes the form \cite{JacobJOpt2014,JacobOptExpress2014}
\begin{equation}
|R_p(\omega)|^2e^{-2qd}<|R_p(\omega_0)|^2e^{-\frac{4\omega_0d}{v}}=1
\label{resid}
\end{equation}
where $\omega_0$ is the frequency of the surface phonon/plasmon polariton at which $R_p(\omega_0)>1$. From (\ref{resid}) follows that the photon emission rate and, consequently, the heat generation and quantum friction diverge for $v>v_c$ where
\begin{equation}
v_c=\frac{2\omega_0 d}{\mathrm{ln}|R_p(\omega_0)|}
\end{equation}
The origin of this divergence is 
associated with the electromagnetic instability when  above the threshold velocity $v_c$ the electromagnetic field increases indefinitely with time  even in the presence of a dissipation in the system  \cite{SilveirinhaNJP2014}. 

For two different plates singular resonance is possible when the real part of the reflection amplitudes  $R_{ip}^{\prime}(\omega_i)=0$ at the frequency of the surface phonon/plasmon polariton $\omega_i$. In this case $\phi_1=\phi_2=\pi/2$ for $q_xv=\omega_1+\omega_2$ and the critical velocity is given by  
\begin{equation}
v_c=\frac{(\omega_1+\omega_2) d}{\mathrm{ln}[R_{1p}^{\prime\prime}(\omega_1)R_{2p}^{\prime\prime}(\omega_2)]^{1/2}}
\end{equation}
where $R_{ip}^{\prime\prime}(\omega)$ is the imaginary part of the reflection amplitude. 

 Near the resonance at $\omega \approx
\omega_1$ and $\omega-q_xv\approx -\omega_2$ for $2/\varepsilon_i^{\prime\prime}>>1$ the reflection amplitudes
 for  dielectrics can be written in the form
\begin{equation}
R_{1p}(\omega)\approx -\frac{a_1}{\omega - \omega_1 +i\Gamma_1},\,\,\,
R_{2p}(\omega-q_xv)=R_{2p}^*(q_xv-\omega)\approx -\frac{a_2}{q_xv-\omega - \omega_2 -i\Gamma_1},
\label{res}
\end{equation}
At the resonance $R_{ip}^{\prime\prime}(\omega_i)=a_i/\Gamma_i=2/\varepsilon_i^{\prime \prime}(\omega_i)$ and the critical velocity
\begin{equation}
v_c=\frac{(\omega_1+\omega_2) d}{\mathrm{ln}2/[\varepsilon_{1}^{\prime\prime}(\omega_1)\varepsilon_{2}^{\prime\prime}(\omega_2)]^{1/2}}
\label{crvelocity}
\end{equation}
For $\omega \approx \omega_2$ and $\omega-
\Omega\approx-\omega_1$  the photon generation rate for  can be written in the form
\begin{equation}
\Gamma_{12}\approx
\frac{4\Gamma_1\Gamma_2a_1a_2e^{-2qd}}
{(\Gamma_1+\Gamma_2)^2(\omega -\omega_c)^2
+\left[\Gamma_1\Gamma_2\left(\frac{(q_xv-\omega_1-\omega_2)}{\Gamma_1+\Gamma_2}
\right)^2
-(\omega -\omega_c)^2+ \frac{(\Omega-\Omega_0)(\Gamma_2-\Gamma_1)(\omega-\omega_c)}
{\Gamma_1+\Gamma_2}+
\Gamma_1\Gamma_2-a_1a_2e^{-2qd}\right]^2}
\label{tantipart}
\end{equation}
where 
\begin{equation}
\omega_c=\frac{\Gamma_1(q_xv-\omega_2)+\Gamma_2\omega_1}{\Gamma_1+\Gamma_2}
\end{equation}
For $v>v_c$  the photon generation rate diverges at
 $\omega=\omega_c$ and
$q_xv =
\Omega^{\pm}$, where
\begin{equation}
\Omega^{\pm}=\omega_1+\omega_2\pm(\Gamma_1+\Gamma_2)\sqrt{\frac{4}{\varepsilon_1(\omega_1)\varepsilon_2(\omega_2)}
e^{-\frac{2(\omega_1+\omega_2)d}{v}}-1}.
\end{equation}
Close to the resonance when
\begin{equation}
\frac{\Gamma_1\Gamma_2}{(\Gamma_1+\Gamma_2)^2}\left|\left(\frac{q_xv - \omega_1-\omega_2}{\Gamma_1+\Gamma_2}\right)^2
+1-\frac{4}{\varepsilon_1(\omega_1)\varepsilon_2(\omega_2)}
e^{-\frac{2(\omega_1+\omega_2)d}{v}}\right|\ll 1,
\end{equation}
using (\ref{tantipart}) in (\ref{approximate}) gives the resonant contributions to  the quantum heat generation rate and quantum friction from the frequency range of the anomalous Doppler effect
\begin{equation}
\left(
\begin{array}{c}
F_x\\
P_1
\end{array} \right)
\approx
\frac{\Gamma_1\Gamma_2c^{3/2}}{\pi d^2(\omega_1+\omega_2)}\left(\begin{array}{c}
\hbar q_c\\
\hbar \omega_1
\end{array} \right)\mathrm{ln}\frac{v_c-v}{v_c}
\label{qfhg}
\end{equation}
where $c=\mathrm{ln}2/[\varepsilon_{1}^{\prime\prime}(\omega_1)\varepsilon_{2}^{\prime\prime}(\omega_2)]^{1/2}$ and $q_c=c/d$.

\begin{figure}[tbp]
\includegraphics[width=0.40\textwidth]{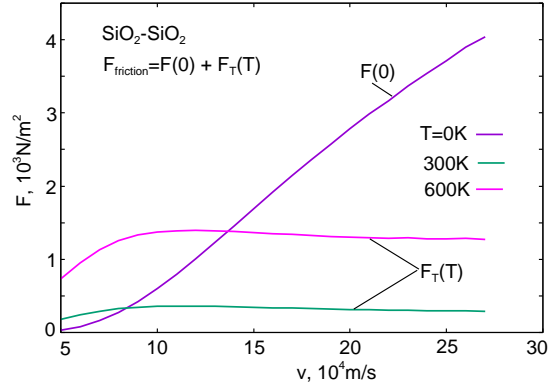}
\caption{The dependence of the different contributions to the friction force  
 on the relative sliding velocity  $v$ between two SiO$_2$ plates  
(see Fig.\ref{Fig1}). The red and green lines show the thermal contributions at $T=600$K and $T=300$K, 
respectively. The blue line shows the quantum contribution ($T=0$K).  The critical velocity $v_c=2.7\cdot 10^5$m/s. The separation 
between the plates $d=1$nm.}
 \label{PlatePlate}
\end{figure}

Fig. \ref{PlatePlate} shows 
the dependence of the friction force, acting between two SiO$_2$ plates sliding relative to each other  with the  relative velocity $v$  at the separation 
$d=1$nm and  for the different  temperatures. The friction force $F=F(0)+
F_T$ where $F(0)$ is the contribution from quantum fluctuations which exist even at 
$T=0$K (this friction is denoted as quantum friction \cite{PendryJPCM1997},  
$F_{friction}(T=0$K)$=F(0))$ and $F_T$ 
is the contribution from the thermal fluctuations which exists only at finite temperatures. 
The thermal contribution dominates for $v<k_BTd/\hbar$ and quantum contribution dominates 
for $v>k_BTd/\hbar$. For SiO$_2$ plates the critical velocity determined by (\ref{crvelocity}) $v_c=2.7\cdot10^5$m/c.  For $v>v_c$ uniform sliding of plates is imposable because the friction force indefinitely increases with time. Close to the critical velocity quantum friction significantly exceeds the thermal contribution to the friction force even for $T=600$K. 

\section{Casimir friction in the tip-plate configuration \label{TP}}

An atomic force microscope tip with the radius of curvature $R\gg d$, 
at a distance $d$ above a flat
sample surface, can be
approximated by a sphere with radius $R$. In this case the friction force
 between the tip and the plane surface can be estimated using the 
approximate method of Derjaguin \cite{Derjaguin1934}, later
called the proximity force  approximation (PFA) \cite{PFA1977}. According to 
this method, the friction force in
the gap between two smooth curved surfaces at short
separation can be calculated approximately as a sum of
forces between pairs of small parallel plates corresponding
to the curved geometry of the gap.  Specifically, the sphere-plane  friction force
 is given by 
\begin{equation}
F=2\pi \int_0^R d\rho \rho f(z(\rho )),  \label{approx}
\end{equation}
where   $z(\rho )=d+R-\sqrt{R^2-\rho^2}$ denotes the tip-surface distance as a function of
the distance $\rho $ from the tip symmetry axis, and the friction force
per unit area $f(z(\rho ))$ is determined in the plate-plate configuration. This
scheme was proposed in \cite{Derjaguin1934,Hartmann} for the calculation of the
conservative van der Waals interaction; in this case the error is
not larger than 5-10\% in an atomic force application, and 25\% in
the worst case situation \cite{Apell}. We assume that the same
scheme is also valid for the calculation of the Casimir friction. However, as it was discussed in Sec. \ref{PP} for two  plates 
the 
Casimir friction diverges at the velocities above the critical velocity $v_c$. However, for  different plates the friction force can be finite 
even above the threshold velocity if at resonance the real part of the reflection amplitude is not exactly zero. For the SiO$_2$ the threshold velocity $v_c\approx 2.7\cdot 10^5$m/s. Thus in the present study the numerical calculations are performed 
at the velocities below the threshold velocity when one can assume that the PFA gives  sufficiently accurate estimation of the Casimir friction 
including the quantum friction.  
During the last few years, the most general method available for calculating
both Casimir force and radiative heat transfer between many bodies of arbitrary shapes, materials,
temperatures and separations was obtained which expresses the
Casimir  force and radiative heat transfer  in terms of the scattering matrices of individual bodies \cite{Dalvit2011}.
Specifically, the numerically exact solution for the
near-field radiative heat  transfer between a sphere and an infinite plane was first performed using the scattering matrix
approach. In principle the same approach can be used for the calculation of the Casimir friction. 
We
assume that the tip has a paraboloid shape given [in cylindrical
coordinates ($z,\rho $)] by the formula: $z=d+\rho ^2/2R$, where
$d$ is the distance between the tip and the flat surface. If
\begin{equation}
f=\frac C{\left( d+\rho ^2/2R\right) ^n},
\end{equation}
we get
\begin{equation}
F=\frac{2\pi R}{n-1}\frac C{d^{n-1}}=\frac{2\pi Rd}{n-1}f(d)\equiv
A_{ \mathrm{eff}}S(d),  \label{approx1}
\end{equation}
where $A_{\mathrm{eff}}=2\pi Rd/(n-1)$ is the effective surface area.
In a more general
case one must use numerical integration to obtain the friction  force.

From 
Eq.(\ref{resfrcoeff}) in the proximity force approximation the friction coefficient in the tip-plate configuration  
\begin{equation}
\Gamma_{res}\approx\frac {\hbar^2\Gamma q_c^4} {8\pi k_BT
 \sinh^2\left(\frac{\hbar \omega_0}{2k_BT}\right)}Rd,
\label{resfrcoeff.tip.plate}
\end{equation} 
 In an 
experiment $\Gamma$ is determined   by measuring  the quality factor
 of the cantilever 
vibration parallel to the substrate surface\cite{MeyerElements2015}. At present can only 
be detected the friction coefficient in the range $10^{-12}-10^{-13}$kg/s. For a   SiO$_2$ tip and a SiO$_2$ plate at $R=1\mu$m  and $d=1$nm   the friction coefficient  
calculated using  Eq. (\ref{resfrcoeff.tip.plate}) is below $10^{-16}$kg/s thus it can not be tested by the modern experimental setup. 
However,  it has been predicted in Ref.\cite{VolokitinPRB2006a}, that for the some 
configurations involving adsorbates the Casimir friction coefficient can be large enough to be measured by state-of-art non-contact force microscope. 

\begin{figure}[tbp]
\includegraphics[width=0.80\textwidth]{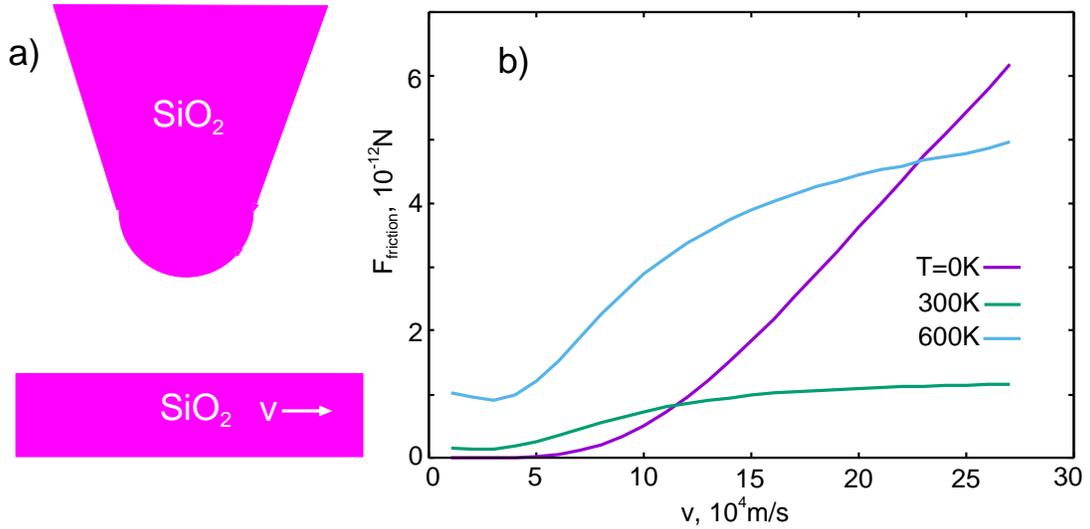}
\caption{(a) A SiO$_2$ tip and a SiO$_2$ plate. The plate  is moving relative to the tip with the  velocity  $v$. (b) The dependence of the different contributions to the friction force  
 on the 
relative sliding velocity  $v$ between a SiO$_2$ tip and a SiO$_2$ plate. The blue and red lines show the thermal contributions at $T=600$K and $T=300$K, 
respectively. The green line shows the quantum contribution ($T=0$K). The radius of the curvature of the tip $R=1\mu$m. The separation 
between the tip and plate $d=1$nm.}
 \label{DD}
\end{figure}

\begin{figure}[tbp]
\includegraphics[width=0.80\textwidth]{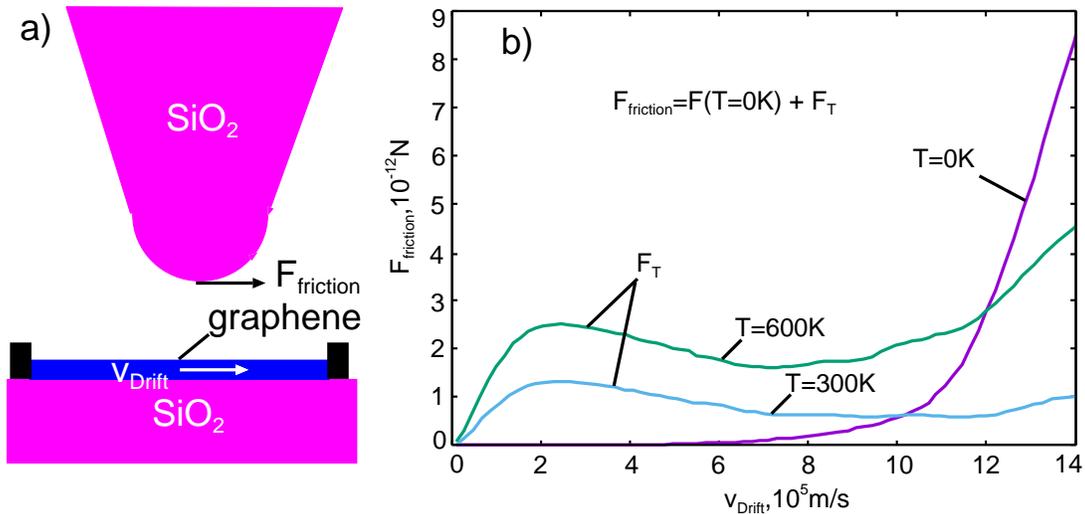}
\caption{(a) A  SiO$_2$ tip and a SiO$_2$ plate covered by   a graphene sheet.  DC current with a drift velocity $v_{drift}$ of  free charge carries is induced in the graphene sheet. (b) The dependence of the different contributions to the friction force  on the drift velocity $v_{drift}$ of the free charge carries in the graphene sheet. The blue and red lines show the thermal contributions at $T=600$K and $T=300$K, 
respectively. The green line shows the quantum contribution ($T=0$K). The radius of the curvature of the tip $R=1\mu$m. The separation 
between the tip and substrate $d=1$nm.}
 \label{DGD}
\end{figure}

During the cantilever vibration the 
velocity 
of the AFM tip does not exceed $1$m/s. However, the Casimir friction force can be strongly 
enhanced at the large relative sliding velocity. This friction force  can be detected 
if it  produces sufficiently large bending of the cantilever. Fig. \ref{DD} shows 
the dependence of the friction force, acting on the SiO$_2$ tip with the radius 
of the curvature $R=1\mu$m, on the relative velocity $v$ between the tip 
and the SiO$_2$ plate  at the separation 
$d=1$nm and  for the different  temperatures. The friction force $F=F_0+
F_T$ where $F_0$ is the contribution from quantum fluctuations which exist even at 
$T=0$K (this friction is denoted as quantum friction \cite{PendryJPCM1997},  
$F_{friction}(T=0$K)$=F_0)$ and $F_T$ 
is the contribution from the thermal fluctuations which exist only at finite temperature. 
The thermal contribution dominates for $v<k_BTd/\hbar$ and quantum contribution dominates 
for $v>k_BTd/\hbar$. On Fig.\ref{DD}  $F>10^{-12}$N at $v>10^5$m/c. In the modern 
experiment\cite{NanoLett2015}
the spring constant of the cantilever are between $k_0=30$ and 
$k_0=50\mu$N/m. The friction force $\approx 10^{-12}$N will produce the displacement 
of the tip of the order $10^2$nm which can be easily detected. However, at present there
is  no  experimental setup with the relative sliding velocity between the tip and 
substrate $\sim 10^5$m/s. An alternative method for the detection of the Casimir friction is possible for the 
SiO$_2$+graphene - SiO$_2$ configuration (see Fig.\ref{DGD}). For this configuration 
inducing current in a graphene sheet with the drift velocity of the free charge 
carriers $v_{Drift}$ will produce 
the fluctuating electromagnetic field which is similar to the electromagnetic field 
due to the mechanical motion of the sheet with the velocity $v=v_{Drift}$ 
\cite{VolokitinJPCM2001b,
VolokitinRMP2007,VolokitinPRL2011,VolokitinEPL2013,VolokitinPRB2016}.  Due to the high mobility of the charge carriers 
in graphene, in a high electric field electrons (or holes)  can move with very high 
velocities (up to $10^6$ m/s). The drift motion  of charge carries in graphene will 
result in a modification of dielectric properties  of graphene  
due to  the Doppler effect \cite{PendryJPCM1997}. Fig. \ref{DGD} shows the friction force  for the SiO$_2$+graphene-SiO$_2$ 
configuration  which is  of the same order of  magnitude as the friction 
force for the SiO$_2$-SiO$_2$ 
configuration (see Fig. \ref{DD}). 

\section{Summary \label{Summary}}
We have studied the Casimir friction, the radiative heat transfer and  heat generation between two plates sliding relative to each other.  We have found that for dynamically nonequilibrium systems in fluctuation-induced electromagnetic phenomena - Casimir friction, radiative heat transfer and heat generation, resonances that do not exist for equilibrium systems are possible.  The origin of these resonances is different in the frequency regions of the normal (NDE) and anomalous (ADE) Doppler effect. While in the region of NDE resonances are associated with the resonant photon tunnelling, in the region of ADE they are associated with the resonant photon emission when excitations are created in both moving media.    In sharp contrast with resonances in the  NDE frequency range, where resonances are finite, in  the  ADE frequency range singular resonances are possible. 
We have discussed the possibility to detect Casimir friction in graphene structures using an atomic force microscope.

\end{document}